\documentclass[aps,pre,twocolumn,showpacs]{revtex4}
\usepackage{graphicx}
\begin{document}
\title{Random global coupling induces synchronization and
nontrivial collective behavior in networks of chaotic maps} 
\author{O. Alvarez-Llamoza$^{1,2}$, K. Tucci$^3$, M. G. Cosenza$^2$, 
and M. Pineda$^4$}
\affiliation{$^1$Departamento de F\'isica, FACYT, Universidad
de Carabobo, Valencia, Venezuela \\
$^2$Centro de F\'isica Fundamental, Universidad de los Andes, Apartado Postal
26, M\'erida 5251, Venezuela\\
$^3$SUMA-CESIMO, Universidad de Los Andes, M\'erida 5251, Venezuela\\
$^4$ Institut fuer Physikalische Chemie und Elektrochemie,
Universitaet Hannover, Germany}
\begin{abstract}
The phenomena of synchronization and nontrivial collective behavior are studied
in a model of coupled chaotic maps with random global coupling. The mean field
of the system is coupled to a fraction of elements randomly chosen at any given
time. It is shown that the reinjection of the mean field to a fraction of
randomly selected elements can induce synchronization and nontrivial collective
behavior in the system.  The regions where these collective states emerge on the
space of parameters of the system are calculated.
\end{abstract}


\maketitle

There is much current interest in the investigation of collective properties of
complex networks of interacting nonlinear elements \cite{Net}. As a model to
study some minimal conditions for the emergence of collective behavior in a
chaotic network, we consider the coupled map system
\begin{equation}
x_{t+1}^i=
\left\lbrace 
\begin{array}{cl}
(1- \epsilon)f(x_t^i) + \epsilon h_t,
& \mbox{with probability} \; p, \\
f(x_t^i),  & \mbox{with probability}  \; 1-p  ,
\end{array}
\right. 
\label{selgloco}
\end{equation}
where $x^i_t$ ($i=1,2,\ldots,N$; $N =$ system size) gives the state of the
$i$th element at discrete time $t$; $\epsilon$ is the coupling strength,
$f$ is a map defining the local dynamics, and
\begin{equation}
\label{Mean}
h_t=\frac{1}{N} \sum_{i=1}^N f(x_t^i)
\end{equation}
is the instantaneous mean field of the system that provides a global coupling.
The parameter $p$ is the probability of connection of an element to the mean
field at time $t$. Thus the average fraction of globally connected elements at
any given time is $p$.

As local dynamics in Eq. (\ref{selgloco}) we shall consider the logarithmic map
$f(x)=b+\log |x|$ \cite{kawabe},
where $b$ is a real parameter and $x \in (-\infty, \infty)$. This map exhibits
robust chaos, with no periodic windows and no separated chaotic bands, in the
parameter interval
$b \in [-1,1]$.

A synchronized state at time $t$ is defined by the condition
$x^i_t=x^j_t, \forall i,j$, in which case the dynamics is described by the
single map $x_{t+1}=f(x_t)$. The synchronization of the elements in the network
can be characterized by the time-average $\langle\sigma\rangle$ of the
instantaneous standard
deviations $\sigma_t$ of the distribution of site variables $x^i_t$, defined as
\begin{equation}
\sigma_t=\left[ \frac{1}{N} \sum_{i=1}^N \left( f(x^i_t) - h_t \right)^2
\right]^{1/2}.
\end{equation}

Figure~1 shows the quantity $\langle\sigma\rangle$ (right vertical axis) as a
function of the probability $p$, for fixed values of $b$ and $\epsilon$. There
is a threshold value $p_c\simeq 0.75$ at which $\langle\sigma\rangle=0$ (within
a precision of $10^{-8}$ in our calculations), indicating that the elements are
synchronized. The range of $p$ where chaotic synchronization takes place is
indicated by the label S. For $p=1$, the system Eq.~(\ref{selgloco}) becomes a
globally coupled map network \cite{Kaneko}
which is known to synchronize. However, Fig.~1 reveals that the reinjection
of a global coupling function to a fraction of randomly selected elements in the
system is enough to achieve synchronization. The critical value $p_c$ for the
emergence of synchronization depends on the parameters of the system, as shown
in Fig.~2.

\begin{figure}[h]
\centerline{
\includegraphics[scale=1.0]{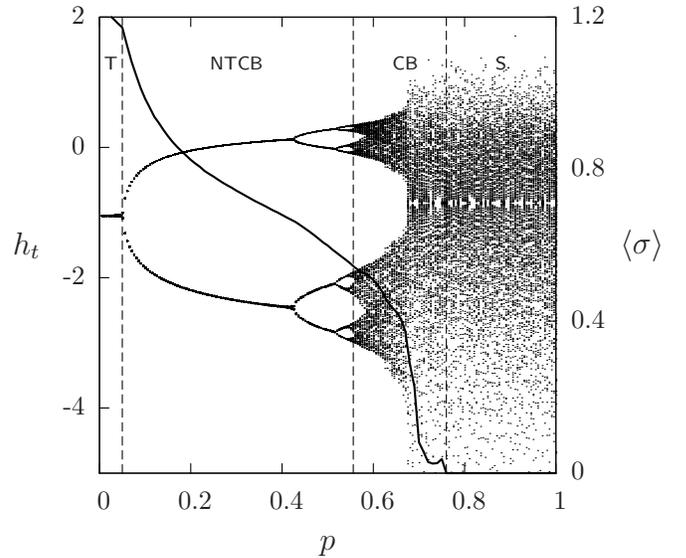}}
\caption{Left vertical axis: bifurcation diagram of $h_t$ as a function of
$p$. For each value of $p$, the mean field was calculated at each time step
during a run starting from random initial conditions on the local maps,
uniformly distributed on the interval $[-8,8]$, after discarding the
transients. The regions where collective states occur are labeled T (turbulent),
NTCB (nontrivial collective behavior), CB (chaotic bands), S (synchronization).
Right vertical axis: $\langle \sigma \rangle$ vs $p$, continuous line. Fixed
$b=-0.7$, $\epsilon=0.4$; size $N=10^5$.}
\end{figure}

The instantaneous mean field of the system $h_t$ can characterize
more complex collective behaviors arising in the system Eq.~(\ref{selgloco}).
When $b \in [-1,1]$, the elements in the network are chaotic and desynchronized.
However, for some parameter values, $h_t$ reveals the existence of global
periodic attractors. Figure~1 shows the bifurcation diagram of $h_t$ (left
vertical axis) as a function of $p$, for fixed $b$ and $\epsilon$. In this
representation, collective periodic states at a given value of the parameter $p$
appear as sets of small vertical segments which correspond to intrinsic
fluctuations of the periodic orbits of the mean field. 

In the region labelled by T (turbulent) in Fig.~1, $h_t$ follows a normal
statistical behavior around a mean value (a collective fixed point) with
fluctuations reflecting the averaging of $N$ completely desynchronized chaotic
elements. Increasing the probability of connection $p$ induces a transition to
periodic collective states occurring in the chaotic range of the local dynamics:
a pitchfork bifurcation takes place from a collective fixed point  to a
collective period-two state (a state for which the time series of $h_t$
alternatingly moves between the corresponding neighborhoods of two separate
values). Collective states of higher periodicity arise by further increasing
$p$: global periodic attractors of period $2$, $4$, $8$, and $16$ are possible
in this system. The amplitude of the collective periodic motions manifested in
$h_t$ do not decrease with an increase in the system size $N$. As a consequence,
the variance of $h_t$ itself does not decay as $N^{-1}$ with increasing $N$ but
rather it saturates at some constant value related to the amplitude of the
collective period. This is a phenomenon of nontrivial collective behavior, where
macroscopic quantities in a spatiotemporal dynamical system exhibit ordered
evolution coexisting with local chaos \cite{Chate}. Note that
the emergence of collective periodic behavior in this system cannot be
attributed to the presence of periodic windows in the local dynamics since the
logarithmic map possesses robust chaos for $b \in [-1,1]$. Figure~1 indicates
with the label NTCB the interval of $p$ where nontrivial collective behavior
arises in this system. Before crossing the boundary of the synchronization
region, the collective states described by $h_t$ take the form of chaotic bands.
These states are labelled CB (collective bands) and they consist of the motion
of chaotic elements that maintain some coherence.

Figure ~2 shows the regions where the different collective states of the system 
Eq.~(\ref{selgloco}) occur on the space of parameters $(p,\epsilon)$. These
regions are separated by stability boundaries which have been numerically
calculated.  For $p=1$, Fig.~2 yields the intervals of stability of the
collective states S, NTCB and CB corresponding to globally coupled logarithmic
maps and which have been previously found \cite{ptp}. Figure~2 shows that those
collective states can also emerge when a fraction of randomly selected maps are
connected to the mean field and appropriate values of the coupling strength are
employed.

\begin{figure}[ht]
\vspace{0.5cm}
\centerline{
\includegraphics[scale=1.0]{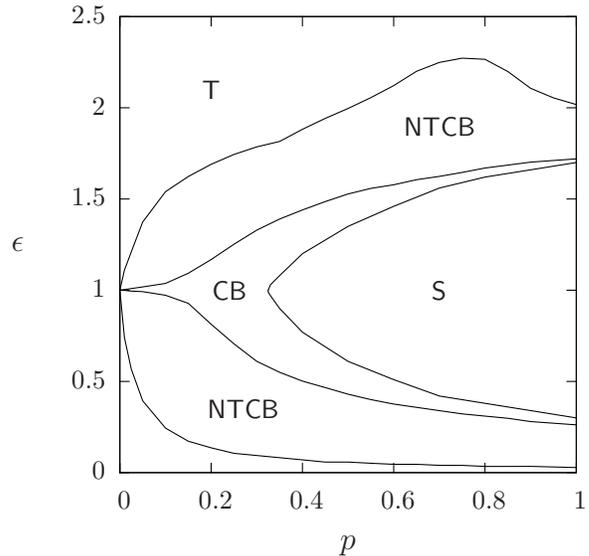}}
\caption{Boundaries on the parameter plane $(p,\epsilon)$ separating different
collective states of the system. Labels correspond to those in Fig.~1. Fixed
$b=-0.7$. Size $N=10^5$.}
\end{figure}

\vspace{0.7cm}

In summary, we have shown that a reinjection of a global coupling function to a
fraction of randomly selected chaotic elements in a dynamical network can induce
synchronization and nontrivial collective behavior in the system.  This
procedure may have practical applications in the control of spatiotemporal
systems. These results may be relevant in some biological and social contexts
where the global information is often available only to a portion of agents in
those systems.

This work was supported by CDCHT, Universidad de Los Andes, under
grant No. C-1396-06-05-B and by FONACIT under grant No. F-2002000426,
Venezuela.

\end{document}